\documentclass[runningheads]{llncs}

\usepackage{tabularx}
\usepackage{booktabs} 
\usepackage{tabulary}
\usepackage[utf8]{inputenc}
\usepackage[T1]{fontenc}
\usepackage{textcomp}
\usepackage{graphicx}
\usepackage{wrapfig} 
\usepackage{subcaption} 
\usepackage{float}%
\usepackage{xcolor}
\usepackage{xcolor,colortbl} 
\usepackage{xspace,tcolorbox}
\usepackage{tikz}
\usetikzlibrary{shapes,arrows,patterns,positioning} 
\usepackage{caption}
\usepackage[normalem]{ulem}
\usepackage{hyperref} 
\usepackage{amssymb}
\usepackage{empheq}
\usepackage[mathscr]{euscript}
\DeclareSymbolFont{rsfs}{U}{rsfs}{m}{n}
\DeclareSymbolFontAlphabet{\mathscrsfs}{rsfs}
\usepackage{mathtools}
\usepackage{stmaryrd} 

\usepackage{dutchcal} 

\usepackage{pgfplots}
\pgfplotsset{compat=newest}

\usepackage[utf8]{inputenc}
\usepackage[T1]{fontenc}
\usepackage{textcomp,mathcomp}

\usepackage{chngcntr} 

\newcommand{\halfcheckmark}{{$\checkmark$}\textsuperscript{\textcolor{black}{\kern-0.53em{\bf--}}}}

\usepackage{threeparttable}

\usepackage{booktabs}
\usepackage{nicematrix}
\usepackage{amsmath}

\usepackage{multirow}
\usepackage[ruled,linesnumbered]{algorithm2e}

\usepackage{breakcites}
\usepackage{enumitem}

\newcommand{\revisit}[1]{{\color{orange}#1}}

\usepackage{xargs} 

\begin{document}

\title{BlowPrint: Blow-Based Multi-Factor Biometrics for Smartphone User Authentication}
\titlerunning{BlowPrint: Blow-Based MFB for Smartphone User Authentication}

\author{Howard Halim \and
Eyasu Getahun Chekole\thanks{Corresponding author.} \and
Dani\"el Reijsbergen \and
Jianying Zhou}
\authorrunning{H. Halim et al.}
%
\institute{Singapore University of Technology and Design, Singapore\\
\email{\{howard\_halim,eyasu\_chekole,daniel\_reijsbergen,jianying\_zhou\}@sutd.edu.sg}}

\maketitle


\vspace{-0.14cm}
\begin{abstract}
%
Biometric authentication is a widely used security mechanism that leverages unique physiological or behavioral characteristics to 
authenticate users. In multi-factor biometrics (MFB), 
multiple biometric 
modalities, e.g.,  physiological and behavioral biometrics, are integrated to mitigate the limitations inherent in single-factor biometric systems. 
The primary research challenge within MFB lies in identifying novel behavioral techniques capable of meeting critical criteria, including high accuracy, high usability,  non-invasiveness, resilience against spoofing and other known attacks, 
and low use of computational resources.
Despite ongoing advancements, current behavioral biometric techniques often fall short of fulfilling one or more of these 
requirements. 
In this work, we propose \emph{BlowPrint}, a novel 
behavioral biometric 
technique 
that allows us to authenticate users based on their phone blowing behaviors. In brief, we assume that the way users blow on a phone screen can produce distinctive acoustic patterns, which can serve as a unique behavioral biometric identifier for effective user identification or authentication. 
The acoustic features of blowing, such as differences in pattern, intensity, frequency, and timing, are unique to each person, making this technique highly accurate, non-invasive, and exceedingly 
robust against 
spoofing and other attacks. 
Moreover, it can be concurrently performed and seamlessly integrated with other physiological techniques, such as facial recognition, thereby enhancing usability. 
To assess BlowPrint's effectiveness, we conduct an empirical study involving 50 participants from whom we collect blow-acoustic and facial feature data in both sitting and standing modes. Subsequently, 
we compute the similarity scores of the blow-acoustic data using various time-series similarity algorithms, 
while we use a pretrained FaceNet-512 model for the facial recognition features. Finally, we combine the 
similarity scores of the two modalities through score-level fusion and compute the accuracy using a machine learning-based classifier. As a result, the proposed method demonstrates an accuracy of 99.35\% for blow acoustics, 99.96\% for facial recognition, and 99.82\% for the combined approach. The experimental results demonstrate BlowPrint's high effectiveness in terms of authentication accuracy, spoofing attack resilience, usability, non-invasiveness, and other aspects. 

\end{abstract}
\vspace{-0.55cm}
\section{Introduction}



The increasing reliance on digital services has necessitated robust authentication mechanisms to protect user data. Traditional password, PIN or key-based authentication systems are vulnerable to a wide range of attacks, including phishing, brute-force, social engineering, and side-channel attacks \cite{chimuco2023secure,angcsur2025}. Biometric authentication, which leverages physiological (e.g., fingerprints, face, iris and retina) or behavioral (e.g., gait, voice and keystroke dynamics) traits, offers a more secure alternative \cite{dargan2020comprehensive}. 
It is increasingly used as a secure and convenient method for identity verification, replacing or complementing traditional password-based systems. However, single-factor biometric authentication methods 
have several limitations and are often insufficient to address the growing sophistication of modern cyberattacks. For example, 
 physiological biometric methods are vulnerable to a wide range of attacks, such as spoofing, forging and deepfaking \cite{wang2024deepfake, madan2023effect}. On the other hand, 
behavioral biometric methods offer resilience against these attacks. However, accuracy and stability remains a major challenge in behavioral biometric due to variations in user behavior, 
environmental sensitivity, and temporal instability, leading to lower accuracy and reliability problems \cite{ayeswarya2024comprehensive}. For instance, voice-based authentication is highly susceptible to background noise, while gait recognition can be influenced by changes in footwear or walking surfaces.  

To alleviate the shortcomings in single-factor biometrics, multi-factor biometrics (MFB) have been increasingly adopted. 
MFB can enhance the security, resilience, and robustness of the authentication process by integrating and utilizing complementary data derived from multiple biometric  
factors, typically through a combination of physiological and behavioral traits. 
The main challenge in MFB lies in the design of novel behavioral biometric techniques that satisfy several critical criteria, including high accuracy, resilience against 
known attacks (e.g., spoofing resistance), usability (e.g., non-intrusive and seamless integration with other modalities), non-invasiveness, 
and minimal computational resource requirements, among others. 
To overcome these challenges, we propose a novel 
behavioral biometric technique 
that effectively authenticates users based on their phone blowing behaviors.
In brief, we hypothesize that the manner in which individuals blow on their phone screen produces distinctive and unobtrusive acoustic patterns, which can serve as a unique behavioral biometric 
for effective user authentication. The blow-acoustic signals are captured in audio waveform by the phone’s built-in microphone and constitute a novel modality 
for user authentication and verification. This novel approach offers a distinctive set of salient features and significant advantages over existing behavioral biometric techniques: 
 \begin{itemize}
    \item \emph{Enhanced accuracy}: 
    The acoustic characteristics of blowing, defined by variations in pattern, intensity, frequency, and timing, is unique to each individual, which renders the modality highly accurate and exceedingly difficult for adversaries to replicate.
    \item \emph{Resistance to known attacks}: Unlike voice-based biometrics, which are susceptible to being recorded and replayed, blow acoustics are inherently less prone to such spoofing or replicating attempts, thereby offering greater security.
    \item \emph{High usability and seamless integration}: This modality can be effortlessly incorporated alongside 
    certain physiological biometric techniques, such as facial recognition, enhancing overall system usability without compromising functionality.
    \item \emph{Non-invasive nature}: The blow-acoustic process is entirely contactless and touch-free, ensuring a non-intrusive user experience that aligns with modern expectations for hygiene, convenience, and user privacy. 
    \item \emph{Rapid execution with minimal resource requirements}: The authentication process is swift, requiring only a few brief blowing samples, and operates solely using the device’s native microphone—eliminating the need for additional hardware.
\end{itemize}
 To establish a robust MFB system, we seamlessly integrate the blow-acoustic behavioral biometric technique with a facial recognition physiological biometric technique, which leverages the uniqueness of facial features for user identification and authentication. The unique patterns in a person's blow-acoustic behavior and facial recognition features can be captured simultaneously using the phone's built-in microphone and camera, respectively, to offer a more accurate and robust user authentication system. 
These modalities are also seamlessly integrated using score-level fusion \cite{rasool2021feature}. 
This
significantly enhances usability, mitigates integration complexity and 
reduces processing times in MFB. 

As a proof-of-concept, we have implemented the BlowPrint application and 
collected time-series data comprising blow-acoustic signals and facial features from 50 participants. Each participant performed 10 sessions in sitting and standing modes. We then evaluated the accuracy of the proposed technique using multiple metrics from the literature, e.g., the false acceptance and rejection rates. To this end, we first compute the similarity score across different blow-acoustic 
patterns collected from various users and sessions. 
This is achieved using a range of widely recognized similarity algorithms, such as Euclidean Distance (ED) \cite{elmore2001euclidean}, Dynamic Time Warping (DTW) \cite{sakoe1978dynamic,dtwR}, Shape Dynamic Time Warping (shapeDTW) \cite{zhao2018shapedtw,shapedtw_repo}, DTW+S, \cite{srivastava2023dtw+,dtws_repo}, Shape-Based Distance (SBD) \cite{paparrizos2015k,dtwclustR}, and Time Warp Edit Distance (TWED) \cite{marteau2008time}. 
The resulting similarity scores are also compared to determine the most effective similarity algorithm for the proposed technique. Furthermore, we employ a pretrained FaceNet-512 model \cite{facenet512android} to compute the similarity scores across different facial features collected from the participants. Finally, we combined the similarity scores of the two techniques through score-level fusion and compute the accuracy using the $k$-Nearest Neighbors ($k$NN) algorithm \cite{kramer2013k}, a machine learning-based classifier. Overall, the blow-acoustic technique achieves an accuracy of 99.35\% and a false acceptance rate of 0.42\% in our dataset, while the facial recognition technique achieved an accuracy of 99.96\%, with a false acceptance rate of 0.04\%. Furthermore, the score-level fusion of both techniques yielded an accuracy of 99.82\%, with a false acceptance rate of 0.18\%.

Overall, this work offers the following main contributions.
\begin{itemize}
    \item We 
    introduce a novel behavioral biometric technique based on phone blowing acoustics, which provides numerous advantages over most existing techniques. This includes high accuracy, resilience against replication and spoofing attacks, enhanced usability, seamless integration with other modalities, 
    non-invasiveness, high robustness in different postural modes, 
    rapid execution, and minimal resource requirements. 
    \item The seamless integration of the proposed phone blowing acoustics technique with facial recognition enhances overall security by complementing the robust security capabilities of the latter, which is grounded in physiological biometrics. 
    \item We conducted a comprehensive evaluation of the proposed MFB technique by developing a prototype application, BlowPrint, and collecting empirical data from 50 participants. The technique demonstrated high effectiveness, achieving an accuracy of 99.59\% for the blow acoustics, 99.96\% for the facial recognition, and 99.82\% for the combined approach 
    through score-level fusion.
\end{itemize}



\section{Related Work}
\label{sec:related_work}

Recent works in biometric authentication systems have explored a variety of modalities, with the main categories being physiological biometrics, behavioral biometrics or a combination of the two. 

\subsection{Physiological Biometrics}


Physiological biometrics, also known as biological biometrics refer to the use of inherent physical characteristics of individuals for identity authentication and recognition. Common physiological biometrics modalities include the face \cite{tolba2006face}, iris\cite{wildes1997iris, daugman2009iris}, fingerprint\cite{maltoni2009handbook}, ear\cite{abaza2013survey}, and hand geometry\cite{sanchez2000biometric}. These biometrics are unique to each individuals, which has led to their long-standing use in authentication system \cite{chaudhari2013historical}

Despite their reliability and widespread adoption, these type of biometrics have several limitations. Some modalities, such as fingerprint scanning, may be considered invasive or raise hygiene concerns. Others like face recognition, are vulnerable to spoofing attacks using photographs. Furthermore, advances in artificial intelligence and machine learning enabled the creation of deepfake attakcs \cite{alrawili2024comprehensive}. Hence, systems that lack robust liveness detection mechanisms are particularly vulnerable to such attacks, emphasizing the need for enhanced security measures in biometric authentication.


\subsection{Behavioral Biometrics}

Behavioral biometrics have gained increasing attention as a non-intrusive and user-friendly means of authentication. These modalities leverage unique patterns in user behavior such as breathing \cite{chauhan2017breathprint}, touch interactions \cite{de2012touch}, and keystroke dynamics \cite{zheng2014you}. 

De Luca et al.\ \cite{de2012touch} proposed a touch-based behavioral biometric authentication system layered on top of the traditional pattern password mechanism in smartphones. The system utilizes the speed, pressure, and rhythm of touch input for user authentication. Similarly, Zheng et al. \cite{zheng2014you} proposed a tapping-based behavioral biometric authentication system for smartphone. It captures the acceleration, pressure, size, and time of the keystroke, and uses a one-class machine learning algorithm for authentication. However, both approaches are inherently invavise, as they require direct physical interaction with the smartphone screen, which may not be suitable for all scenarios or user preferences.
Chauhan et al. \cite{chauhan2017breathprint} analyzes breathing patterns to identify users based on three types of breathing behavior: sniffing, normal, and deep breathing. The study demonstrates that breathing behavior is unique, achieving a true positive rate (TPR) as high as 94'\%'. However, its evaluation is limited to only these 3 breathing patterns, raising questions about its scalability and adaptability in addition to the low accuracy rate achieved.

Despite the promise of unimodal behavioral biometrics, they suffer from several inherent limitations. These include reduced robustness, as accuracy can degrade in the presence of noise, environmental variations, or changes in user behavior, and increased susceptibility to spoofing or imitation attacks \cite{sumalatha2024comprehensive, tolosana2020deepfakes, galbally2014biometric}. 


\subsection{Multi-Factor Biometrics}

Recent studies have explored the combination of multiple biometric modalities to improve authentication accuracy. Multi-factor biometric authentication systems can offer enhanced security and robustness by leveraging complementary information from different biometric traits. Most existing multimodal biometric systems predominantly rely on physiological traits, such as facial features, irises, touch, and fingerprint. 

Al-Wasy et al. \cite{al2017multimodal} proposed a multimodal biometric system using a deep learning approach that integrates facial features and both left and right irises, employing a fusion module at the score and rank levels. Aizi et al.\cite{aizi2022score} implemented a score-level fusion strategy based on fingerprint and iris as its multimodal biometric. Srivastava et al. \cite{srivastava2022match} also proposed a multimodal biometric system combining finger-knuckle print and iris data, then applying a neuro-fuzzy classifier at the match level. However, these proposed system exclusively relies on physiological biometrics, without incorporating behavioral modalities, limiting its adaptability in scenarios where physical traits may be unavailable or compromised. Moreover, \cite{aizi2022score, srivastava2022match} biometrics modalities do not share the same anatomical region which can be awkward in practical use \cite{koffi2023voice}, and with the absence of behavioral biometrics may further reduce resilience against sophisticated attacks.

Mahfouz et al. \cite{mahfouz2024m2auth}, a multimodal behavioral biometric authentication system that leverages feature-level fusion of various behavioral modalities, including touch gestures, dynamic keystroke, and accelerometer data. While it reduces reliance on physical traits, systems that depend solely on behavioral biometrics may suffer from variability due to a user’s mood, health, or environmental distractions, potentially affecting accuracy, robustness and usability.

El Rahman et al. \cite{el_rahman2020multimodal} proposed a hybrid approach combining ECG and fingerprint biometrics using multiple fusion strategies. While the combination of physiological and behavioral modalities aims to enhance security and robustness, the system remains somewhat invasive, requiring users to directly scan the fingerprint while additional equipment outside smartphone is required to capture the ECG signals.

Lee et al. \cite{lee2021advanced} developed an authentication method for IoT devices that processes both touch and motion data using sensors from a smartphone and smartwatch. While effective in concurrent data acquisition, the method requires users to wear a smartwatch, limiting convenience and applicability in scenarios where such smartwatches are unavailable.

Wu et al. \cite{wu2022echohand} proposed an authentication system that utilizes hand geometry features and acoustic sensing technique. While Zhou et al. \cite{zhou2018echoprint} proposed an authentication system that utilizes facial landmarks and acoustic sensing. Both system utilizes echoes as an acoustic features. However, the accuracy and equal error rate (EER) presented in these systems are not as high as those achieved by our approach, as shown in Section \ref{sec:accuracy_evaluation}


Despite these advancements, existing multimodal systems still face notable limitation. As highlighted by Koffi et al \cite{koffi2023voice}, among the various biometrics modality combinations, voice and face biometrics stand out for their balance of robustness, security, and usability. Particularly due to their inherent support for liveness detection and minimal of intrusiveness. Building upon these insights, our proposed implementation addresses the aforementioned challenges by introducing a novel fusion of physiological and behavioral modalities that ensures improved security, usability, while maintaining high accuracy. A qualitative and quantitative comparison of the relevant related works and that of our proposed technique is presented in Table \ref{table_comp_related_works}.

\section{Threat Model and System Requirements}

\subsection{Threat Model}

In our threat model, we consider various types of attacks that specifically target 
biometric systems. 
In particular, 
we assume the following capabilities of the adversary. 
\begin{itemize}
    \item Sensors are not compromised by the adversary and can produce the expected biometric data.
    \item 
    Biometric data (i.e., blow-acoustic and facial image data) is stored securely, and can not be read or manipulated by the adversary.
    \item Spoofing attacks may use synthetic biometric samples (e.g., deepfake face images, recorded blow sound, and replicated blow-acoustic patterns) to bypass authentication.
    \item Biometric duplication attacks may exploit residual physiological biometrics such as facial images publicly available online.
    \item Replay attacks can use previously recorded blow sounds or static images to spoof the system, but the current blow pattern cannot be known to the adversary.
\end{itemize}


\subsection{System Requirements}
\label{sec:requirements}
In the following, we outline the key aspects that we consider as essential requirements within the context of behavioral or multi-factor biometrics. These requirements also serve as our evaluation criteria to evaluate both existing and proposed biometric techniques. 

\subsubsection{Accuracy}

A behavioral biometric technique should demonstrate high accuracy to effectively minimize the risk of impersonation attacks (i.e., by having a low false positive rate), while also enhancing user convenience by lowering the authentication attempts (i.e., by having a low false negatives). 
\subsubsection{Resilience Against Known Attacks}
A behavioral biometric technique should demonstrate high resilience against certain known attacks, including the following:
\begin{itemize}
\item \emph{Spoofing attacks}: Attacks that may use synthetic biometric samples (e.g., deepfake face images, recorded voice) to bypass authentication.
\item \emph{Replication attacks}: Attacks that attempt to replicate biometric traits through brute-forcing or other techniques. 
\item \emph{Privacy leakage}: Unauthorized data collection and misuse of biometric information.
\end{itemize}

\subsubsection{Usability}
One of the primary challenges in behavioral biometrics and MFB is usability. In this regard, the key usability factors include:
\begin{itemize}
\item \emph{User convenience}: The system should be intuitive and non-intrusive.
\item \emph{Response time}: Authentication should be fast and seamless.
\item \emph{Seamless integration}: 
Combining multiple biometric techniques in a single authentication attempt (one-shot authentication) improves user experience by minimizing the need for separate actions and reducing processing time. For instance, face and voice recognition 
can be conducted concurrently, thereby eliminating the requirement for sequential authentication steps. 
\end{itemize}

\subsubsection{Non-Invasiveness}
 To address hygiene-related concerns, behavioral biometric techniques should be performed in a touchless manner. Moreover, data collection should be limited to brief durations (typically no more than a few seconds) and must not occur without the user's awareness that their behavior is being monitored.
 

 \subsubsection{MFB Support}

In light of the inherent limitations of single-factor biometric systems, it is imperative that biometric authentication techniques incorporate multiple biometric modalities, typically through the integration of both behavioral and physiological factors.
 
\subsubsection{Low Resource Requirements}
Behavioral biometric authentication should be conducted using minimal resources, i.e., the computational resources and sensors available on a regular smartphone, without requiring additional hardware or software tools.
\section{BlowPrint: Proposed Technique}

\subsection{Overview}
\begin{wrapfigure}{r}{0.45\textwidth}
    \vspace{-1cm}
    \centering  
    \includegraphics[width=0.45\textwidth]{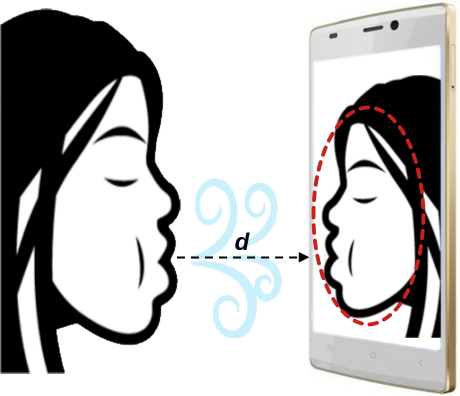}  
    \caption{Illustration of BlowPrint }
    \label{fig_blowprint_architecture}
    \vspace{-0.5cm}
\end{wrapfigure}

This section presents a detailed description of \emph{BlowPrint}, a novel behavioral biometric technique that effectively authenticates users based on their phone blowing behaviors. 
It is also seamlessly integrated with a facial recognition physiological biometric technique to form a robust and effective MFB, Using the BlowPrint application, the phone blowing acoustic signals are recorded using the phone's built-in microphone, while facial features are captured using the front camera of the phone. The phone is positioned at a fixed distance $d$ to ensure uniformity in blow sound and facial image captures. This is achieved by letting the user position her face inside an oval shape indicator. The process is illustrated in Figure \ref{fig_blowprint_architecture}. 




\subsection{Workflow}
\begin{figure*}[htb]
    \centering  
     \includegraphics[scale=0.42]{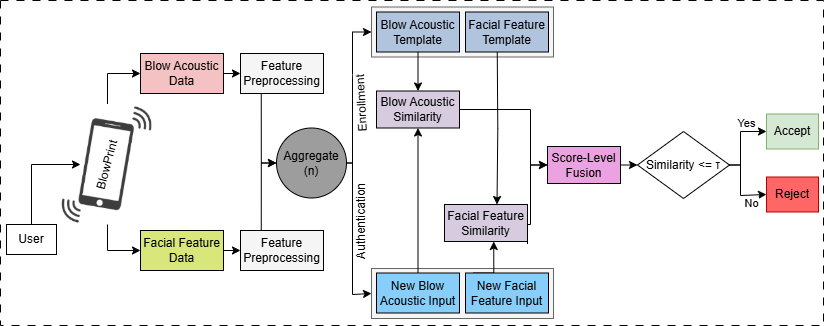} 
    \caption{A high-level workflow of BlowPrint}
    \label{fig_blowprint_work_flow}
\end{figure*}
This section outlines the workflow of BlowPrint, detailing the main activities and phases involved in the proposed authentication procedure. A high-level architecture of the workflow is illustrated in Figure \ref{fig_blowprint_work_flow}. 
The process begins with users interacting with the BlowPrint application on a running mobile device, which 
concurrently captures the blow-acoustic and facial feature data. 

\subsubsection{Data Collection}\label{sec_data_collection}
To evaluate the effectiveness of the proposed biometric technique, the blow-acoustic and facial feature data were collected from several users using the BlowPrint application. 
A standard smartphone model with a 
built-in microphone and 
front-facing camera were used to collect the blow-acoustic and facial feature data, respectively. Since the intensity and consistency of the blow-acoustic data as well as quality of the face image can be affected by the distance between the user and the phone, we set a fixed distance $d$ between the user and the phone while performing data collection. This is achieved by designing an appropriate oval-shaped indicator in the application where the users need to positioned their face in it before data collection is activated. When the position is validated, the camera captures 
facial features and and the microphone records blow-acoustic signals. 
Users are required to blow into the microphone while facing the camera, ensuring that both modalities are captured in one shot. 

\subsubsection{Feature Preprocessing}

These raw biometrics data undergo dedicated pre-processing stages to extract suitable features for matching. In the blow-acoustic modality, the system captures the raw audio signal (amplitude) at a sampling rate of 48kHz, recording every 0.02 seconds, which yields 960 samples per window. Each sample window is reduced to a single value using the Root Mean Square (RMS) operation to represent its intensity. The resulting 5-second RMS signal is then refined using a Simple Moving Average (SMA) filter \cite{macaulay1931introduction} to suppress noise and smooth short-term fluctuations.

For the facial modality, the application utilizes the Google ML Kit Android Face Detection Library \cite{googlemlkitface} to detect and crop facial regions. The cropped facial images are subsequently passed through a deep learning model which generates a facial embeddings used for the facial similarity computation

Following feature preprocessing, the systems proceed with a parallel biometric authentication pipeline for blow-acoustic and facial recognition modalities, each performing similarity computation before being combined via score-level fusion and compared against the threshold ($\tau$).



\subsection{Similarity Computation}
\label{sec:calculation}

\subsubsection{Blow-Acoustic Similarity Computation}

The similarity computation for the blow-acoustic modality was conducted by leveraging a series of time-series similarity algorithms. These algorithms serve as the foundation for measuring the similarity between enrolled and query signals, which are then compared using a decision threshold to determine whether an authentication attempt is accepted or rejected. This similarity-based decision process forms the basis for evaluating the overall accuracy of the system. The algorithms employed include the following.
\begin{itemize}
    \item 
    \textbf{ED} \cite{elmore2001euclidean}: A point-to-point similarity measure used to compute distance between 2 time-series by directly comparing corresponding elements.
    \item 
    \textbf{DTW} \cite{sakoe1978dynamic,dtwR}: A time-series similarity measure that allows non-linear alignments by compares each point to the closest matching point in the other sequence while preserving temporal order.
    \item 
    \textbf{shapeDTW} \cite{zhao2018shapedtw,shapedtw_repo}: An extension of the DTW algorithm that incorporates local shape descriptors to align segments with similar structural patterns. In this experiment, the compound shape descriptor, combining raw values and their first-order derivatives, was used to enhance local structure matching.
    \item \textbf{DTW+S} \cite{srivastava2023dtw+,dtws_repo}: A DTW-based technique that captures similarity using a shapelet representation matrix, enhancing alignment through discriminative local patterns.
    \item 
    \textbf{SBD} \cite{paparrizos2015k,dtwclustR}: A similarity measure based on normalized cross-correlation that aligns time-series based on the most correlated subsequences.
    \item 
    \textbf{TWED} \cite{marteau2008time}: A time-series similarity measure that combines edit distance with time-lag penalties, allowing flexible alignment by accounting the temporal distortions and magnitude differences.
\end{itemize}

\subsubsection{Facial Similarity Computation}
Various deep learning models have been developed for facial recognition, including FaceNet \cite{schroff2015facenet}, VGG-Face \cite{parkhi2015deep}, DeepFace \cite{taigman2014deepface}, ArcFace \cite{deng2019arcface}, and others. While each model has its own characteristics, such as different embedding sizes, loss functions, and backbone architectures, Serengil et al. \cite{serengil2024benchmark} provide a comprehensive benchmark comparing the performance of these models on the same dataset. The results show that FaceNet-512 consistently achieves the highest accuracy among the evaluated models.

Based on these findings, the similarity computation for the facial modality in our system was conducted using a pretrained FaceNet-512 model \cite{facenet512android}. This model maps cropped facial images into 512-dimensional embeddings, where facial similarity is computed using cosine similarity \cite{steck2024cosine} between the embeddings of enrolled and query images.

\subsubsection{Aggregated Similarity Computation}
To compute the overall accuracy of the proposed technique, we aggregate the 
similarity scores of the two modalities using the score-level fusion technique \cite{rasool2021feature}. 
Although various fusion methods are available, we employ score-level fusion in our approach, as it demonstrates superior performance compared to alternative techniques. In this method, the matching scores derived from both blow-acoustic and facial recognition modalities are first normalized using the min-max normalization technique \cite{min_max_normalization} and subsequently combined using the weighted summation method with equal weights \cite{weighted_summation}. The resulting fused score is subsequently evaluated using the $k$NN algorithm. For a given value of $k$, the $k$ closest similarities to the enrolled templates are identified and compared against a predefined threshold, $\tau$. If the similarity falls below $\tau$, the user is authenticated; otherwise, access is denied. 
For each user, $\tau$ is dynamically determined according to the $k$ and targeted 
recall value $q$, as outlined in Section \ref{sec:accuracy_evaluation}.

\subsection{Enrollment and Authentication Phases}

As in any conventional biometric systems, the proposed biometric technique involves the usual enrollment and authentication phases. During the enrollment phase, a biometric template is generated for each user based on preprocessed and fused blow-acoustic and facial feature data collected over multiple sessions. This template is securely stored and serves as the user's unique biometric identifier. During the authentication phase, the system receives new blow-acoustic and facial feature inputs from a user. Like in the enrollment phase, these inputs undergo the preprocessing stage before computing the similarity scores of each modalities by comparing the resulting data with the enrolled biometric template. These scores are then combined using score-level fusion. If the fusion similarity score falls within a predefined threshold $\tau$, i.e., $Similarity \leq \tau$, the user is successfully authenticated; otherwise, it is rejected.

\subsection{Implementation Details}

To validate the proposed technique, a proof-of-concept application, BlowPrint, was implemented on the Android platform. This application allows the capture of users' phone blowing acoustic signals and facial features for both the enrollment and authentication phases. It leverages Android's native APIs for real-time media processing. Specifically, audio signals are captured using the \texttt{android.media} package \cite{androidmedia}, which provides access to the raw audio recording. For facial images, the \texttt{android.hardware.camera2} package \cite{androidcamera2} is used, along with Google ML Kit Android Face Detection Library \cite{googlemlkitface} to identify and crop suitable regions for facial preprocessing.

Furthermore, an automated evaluation framework using R and Python was implemented, enabling the end-to-end processing pipeline, from data collection to performance evaluation, to be conducted seamlessly. The main evaluation logic is developed in R, with support from several Python libraries integrated via the \texttt{reticulate} package \cite{reticulate}, which allows calling Python code directly within the R environment. The evaluation framework comprises a set of modules that utilize the similarity algorithms discussed in Section \ref{sec:calculation} to compute 
the accuracy of each biometric modality and their fusion.





\section{Evaluation and Discussion}

In this section, we evaluate the effectiveness of BlowPrint through an empirical study involving acoustic and facial feature data from human participants. We investigate the accuracy achieved by the various similarity computation techniques from Section~\ref{sec:calculation} and use the score-level fusion method  to combine the blow-acoustic and facial feature datasets. 


\subsection{Data Collection and Extraction}


We conducted an empirical study consisting of 50 participants, including 40 males and 10 females. The participants were randomly chosen from various
demographic groups with an approximate age range of 22 to 65 years. 
Each participant performed 10 sessions which each lasted 5 seconds. 
To ensure robustness of the proposed technique in different postures and contexts, two types of data collection modes -- sitting and standing -- were used: each participant performed 5
sessions while sitting and 5
sessions while standing. 
The collected blow-acoustic signals and facial features were stored in CSV format for further processing.

\begin{figure*}
     \centering
     \begin{subfigure}{0.49\textwidth}
         \centering
         \includegraphics[width=\linewidth]{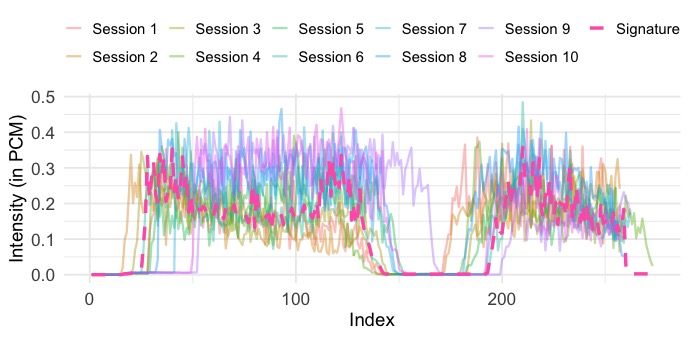}
         \caption{}
         \label{fig_participant1}
     \end{subfigure}
     \begin{subfigure}{0.49\textwidth}
         \centering
         \includegraphics[width=\linewidth]{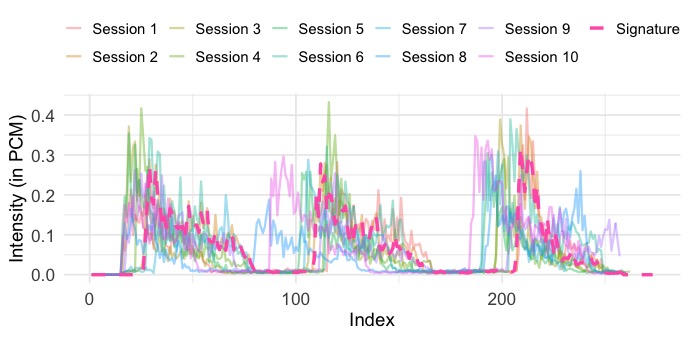}
         \caption{}
         \label{fig_participant2}
     \end{subfigure}
      \begin{subfigure}{0.49\textwidth}
         \centering
         \includegraphics[width=\linewidth]{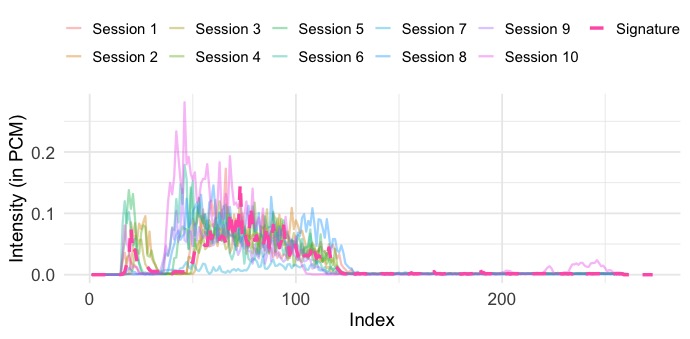}
         \caption{}
         \label{fig_participant3}
     \end{subfigure}
     \begin{subfigure}{0.49\textwidth}
         \centering
         \includegraphics[width=\linewidth]{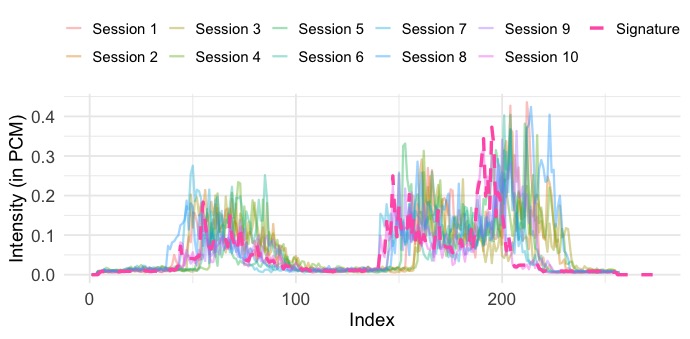}
         \caption{}
         \label{fig_participant4}
     \end{subfigure}
     \caption{Sample raw blow-acoustic data of (a) Participant 1 (b) Participant 2 (c) Participant 3 (d) Participant 4, where ``Signature'' is the DBA-based aggregated data of the 10 sessions}
     \label{fig_sample_raw_acoustic_data}
\end{figure*}

For illustration purposes, we depict the raw blow-acoustic data of four participants  in Figure \ref{fig_sample_raw_acoustic_data}, demonstrating different blow patterns  of different users performed in 10 sessions for a span of 5 seconds each. The raw blow-acoustic data was captured using the \texttt{android.media} library\cite{androidmedia}, specifically the \texttt{AudioRecord} class, which leverages the \texttt{ENCODING\_PCM\_FLOAT} format to represent the audio intensity. In the figures, the red dotted line (i.e., ``Signature'') represents an aggregated data of the 10 sessions, generated using the DBA algorithm \cite{Petitjean2011_DBA} -- this is for illustration purposes only and not for authentication. 
    
\begin{figure*}
     \centering
     \begin{subfigure}{0.49\textwidth}
         \centering
         \includegraphics[width=\linewidth]{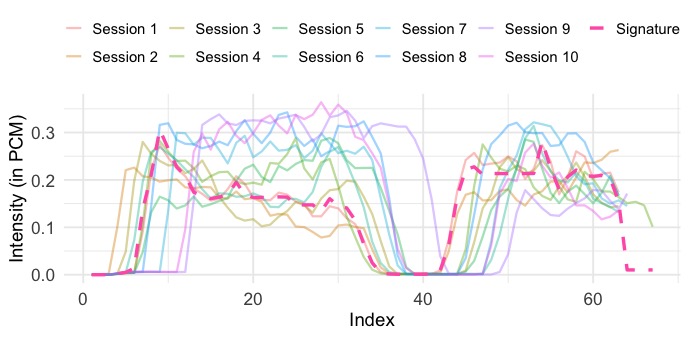}
         \caption{}
         \label{fig_participant1_r}
     \end{subfigure}
     \begin{subfigure}{0.49\textwidth}
         \centering
         \includegraphics[width=\linewidth]{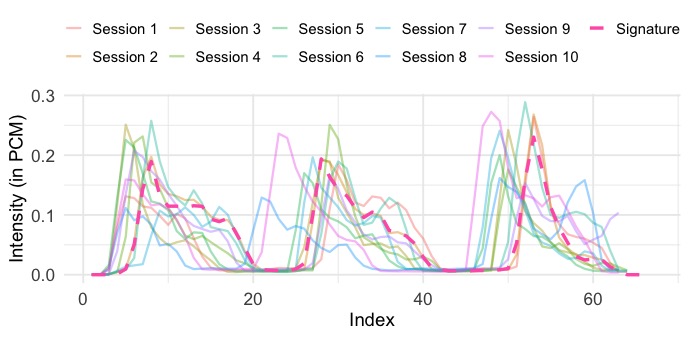}
         \caption{}
         \label{fig_participant2_r}
     \end{subfigure}
      \begin{subfigure}{0.49\textwidth}
         \centering
         \includegraphics[width=\linewidth]{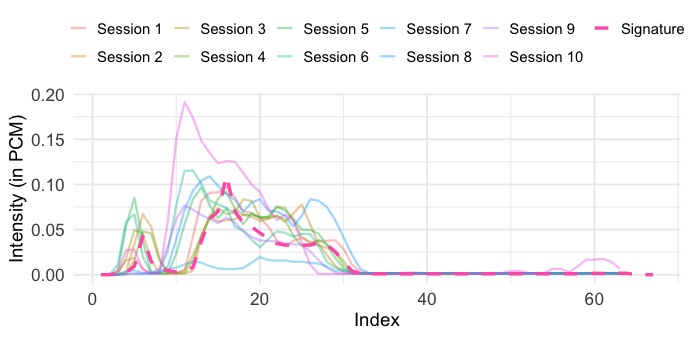}
         \caption{}
         \label{fig_participant3_3}
     \end{subfigure}
     \begin{subfigure}{0.49\textwidth}
         \centering
         \includegraphics[width=\linewidth]{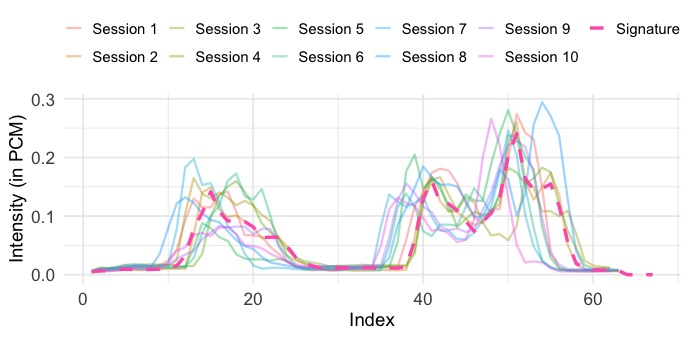}
         \caption{}
         \label{fig_participant4_r}
     \end{subfigure}
     \caption{Sample refined blow-acoustic data of (a) Participant 1 (b) Participant 2 (c) Participant 3 (d) Participant 4, where "Signature" is the DBA-based aggregated data of the 10 sessions}
     \label{fig_sample_refined_acoustic_data}
\end{figure*}

The collected raw acoustic data are further refined to suppress noise and other short-term fluctuations. This data refinement is carried out using the Moving Average technique \cite{macaulay1931introduction}, specifically the Simple Moving Average (SMA) with a window size of 8 time slots. For illustration purposes, the refined acoustic data of the four participants are depicted in Figure \ref{fig_sample_refined_acoustic_data}. The 
dataset collected for our experimental evaluation can be found online \cite{blowprint_dataset} (participants' facial images are excluded due to privacy reason).

\subsection{Accuracy Evaluation}\label{sec:accuracy_evaluation}

\subsubsection{Evaluation Methodology}\label{sec:accuracy_evaluation_methodology}

We evaluate the performance of BlowPrint in terms of its resilience to an attacker who has gained access to the device running BlowPrint, but who has no knowledge of the user's blow-acoustic pattern. We use the other users' blow patterns as proxies for the pattern produced by an attacker: successful attempts are recorded as a true positive (TP), and as a false negative (FN) otherwise. Next, we test all of the 490 sessions by other users in the dataset against each user's base readings -- if the user is authenticated, then this is recorded as a false positive (FP) and as a true negative (TN) otherwise. 

Our main accuracy metric is the false acceptance rate (FAR), which is defined as $\frac{\text{FP}}{\text{FP}+\text{TN}}$. Similarly, the false rejection rate (FRR) is defined as $\frac{\text{FN}}{\text{FN}+\text{TP}}$. 
We define the overall accuracy as $\frac{\text{TP}+\text{TN}}{\text{TP}+\text{TN}+\text{FP}+\text{FN}}$. Finally, we consider the effective error rate (EER), which is the minimal achievable maximum of the FAR and FRR. We do note that there is an inherent asymmetry in our setting between a false positive (in which the attacker gains unauthorized access) and a false negative (in which the user needs to retry) because of the relatively low cost of retrying a blow attempt after a failed login. For example, if we consider the following two settings (both from Table~\ref{tab:algorithm_results}), 1) FPR=1.81\%, FNR=0\% and 2) FPR=0.27\% and FNR=20\%, then the first setting has more than a 10$\times$ lower maximum error rate than the second (1.81\% vs.\ 20\%), whereas the second has an FPR that is 6.7\% lower than the first. Since the user may prefer the second setting if she values security over convenience, the ``best'' achievable maximum error rate is not necessarily the most appropriate metric in our setting. As such, we focus on the FAR over the EER as our main accuracy metric.

To compute the accuracy of BlowPrint, we use a range of $k$ values (e.g., $k = 1$ to $k = 4$) and target recall values (e.g., $q = 10$, $q = 9$ and $q = 8$). The authentication threshold $\tau$ for each user is dynamically determined based on the specified $k$ value and target recall value $q$ such that $q$ of her total $n$ sessions would result in successful authentication. 
Higher values of $q$ lead to higher thresholds, which makes it more likely that the user succeeds in authentication, but also easier for the attacker to succeed. Having set a threshold, we compute and compare BlowPrint's accuracy using multiple baseline and state-of-the-art 
similarity algorithms from the literature, including the ED, DTW, shapeDTW, DTW+S, SBD, and TWED as discussed in Section~\ref{sec:calculation}.


\subsubsection{Evaluation Results}
\label{sec:evaluation_results}

The observed EER, accuracy, FAR, and FRR as described previously are displayed in Table~\ref{tab:algorithm_results}. We evaluate three distinct scenarios: using only the sitting data, only the standing data, and the combined dataset. As highlighted above, we also consider 
different values of $k$ (e.g., $k = 1, 2, 3, 4$) and $q$ (e.g., $q = 10, 9, 8$) to compute the accuracy of BlowPrint. The authentication threshold $\tau$ is automatically determined based on these parameters. We observe that the accuracy remains relatively stable for $k$ values ranging from 1 to 4, exhibiting only minor fluctuations between 0.01\% and 0.2\% across different $q$ values. 
Due to space limitation, we present below only the accuracy of BlowPrint for $k=1$ across different $q$ values, computed using the different similarity algorithms mentioned above. Among the different algorithms, DTW demonstrates the highest accuracy in most cases. The lowest FAR in our experiments is achieved by DTW with a target recall of $8$ out of $10$ sessions. 
Under this configuration, users 
may be required to re-record 20\% of login attempts, but with a lower FAR (0.27\%) as a result. 

We observe that the accuracy tends to be higher for data collected in the standing mode than 
in the sitting mode. This discrepancy may be attributed to the fact that participants recorded the sitting data prior to the standing data, thereby gaining familiarity and improved consistency during the latter sessions. Nonetheless, the observed difference in accuracy is relatively minor (smaller than the variations observed across different time series similarity methods), demonstrating the robustness of the proposed technique across different postural modes. Table~\ref{tab:fusion_results} presents the observed EER, accuracy, FAR, and FRR after combining the best blow-acoustic and facial recognition techniques using score-level fusion. Notably, this fusion approach yields a FAR of 0, indicating that no false positives were detected in our experimental evaluation.


\begin{table}[htb]
\vspace{-0.3cm}
\caption{Observed EER, accuracy, FAR, and FRR for a variety of time series analysis techniques, and for various target recall values. Bold values indicate the best FAR results in each category.}
\label{tab:algorithm_results}
\centering
\scalebox{0.85}{
\begin{tabular}{cl|c|ccc|ccc|ccc}
\multirow{2}{*}{Series}   & \multicolumn{1}{c|}{\multirow{2}{*}{Mode}} & \multicolumn{1}{c|}{\multirow{2}{*}{EER}} &\multicolumn{3}{c|}{\begin{tabular}[c]{@{}c@{}}q = 5 (Sit \& Stand)\\ q = 10 (Both)\end{tabular}} & \multicolumn{3}{c|}{\begin{tabular}[c]{@{}c@{}}q = 4 (Sit \& Stand)\\ q = 9 (Both)\end{tabular}} & \multicolumn{3}{c}{q = 8 (Both)}                                  \\
                          & \multicolumn{1}{c|}{}                      & & accuracy                     & FAR                 & FRR                           & accuracy                & FAR                                & FRR               & accuracy        & FAR                        & FRR \\ \hline
\multirow{3}{*}{ED} &
  Sit &
  0.1115 &
  0.8907 &
  0.1115 &
  0.000 &
  0.9498 &
  0.0476 &
  0.1760 &
  - &
  - &
  - \\
 &
  Stand &
  0.0754 &
  0.9261 &
  0.0754 &
  0.000 &
  0.9729 &
  0.0240 &
  0.1800 &
  - &
  - &
  - \\
 &
  Both &
  0.0920 &
  0.8748 &
  0.1278 &
  0.000 &
  0.9521 &
  0.0470 &
  0.0920 &
  0.9726 &
  0.0241 &
  0.1880 \\ \hline
\multirow{3}{*}{DTW} &
  Sit &
  0.0178 &
  {0.9826} &
  \textbf{0.0178} &
  0.000 &
  0.9934 &
  \textbf{0.0029} &
  0.1840 &
  - &
  - &
  - \\
 &
  Stand &
  0.0042 &
  {0.9959} &
  \textbf{0.0042} &
  0.000 &
  0.9948 &
  \textbf{0.0016} &
  0.1840 &
  - &
  - &
  - \\
 &
  Both &
  0.0181 &
  {0.9822} &
  \textbf{0.0181} &
  0.000 &
  0.9924 &
  \textbf{0.0058} &
  0.1000 &
  0.9935 &
  \textbf{0.0027} &
  {0.1920} \\ \hline
\multirow{3}{*}{shapeDTW} &
  Sit &
  0.0274 &
  0.9731 &
  0.0274 &
  0.000 &
  0.9894 &
  0.0073 &
  0.1760 &
  - &
  - &
  - \\
 &
  Stand &
  0.0155 &
  0.9848 &
  0.0155 &
  0.000 &
  0.9900 &
  0.0066 &
  0.1760 &
  - &
  - &
  - \\
 &
  Both &
  0.0370 &
  0.9638 &
  0.0370 &
  0.000 &
  0.9825 &
  0.0160 &
  0.0940 &
  0.9886 &
  0.0077 &
  0.1900 \\ \hline
\multirow{3}{*}{DTW+S} &
  Sit &
  0.0207 &
  0.9798 &
  0.0207 &
  0.000 &
  0.9882 &
  0.0082 &
  0.1880 &
  - &
  - &
  - \\
 &
  Stand &
  0.0165 &
  0.9838 &
  0.0165 &
  0.000 &
  0.9872 &
  \multicolumn{1}{l}{0.0095} &
  0.1760 &
  - &
  - &
  - \\
 &
  Both &
  0.0334 &
  0.9672 &
  0.0334 &
  0.000 &
  0.9862 &
  \multicolumn{1}{l}{0.0121} &
  0.0960 &
  0.9885 &
  \multicolumn{1}{l}{0.0079} &
  0.1860 \\ \hline
\multirow{3}{*}{SBD} &
  Sit &
  0.0504 &
  0.9506 &
  0.0504 &
  0.000 &
  0.9883 &
  \multicolumn{1}{l}{0.0082} &
  0.1840 &
  - &
  - &
  - \\
 &
  Stand &
  0.0224 &
  0.9780 &
  0.0224 &
  0.000 &
  0.9886 &
  \multicolumn{1}{l}{0.0081} &
  0.1720 &
  - &
  - &
  - \\
 &
  Both &
  0.0384 &
  0.9624 &
  0.0384 &
  0.000 &
  0.9849 &
  \multicolumn{1}{l}{0.0138} &
  0.0980 &
  0.9881 &
  \multicolumn{1}{l}{0.0082} &
  0.1900 \\ \hline
\multirow{3}{*}{TWED} &
  Sit &
  0.0752 &
  0.9263 &
  0.0752 &
  0.000 &
  0.9661 &
  \multicolumn{1}{l}{0.0311} &
  0.1720 &
  - &
  - &
  - \\
 &
  Stand &
  0.0502 &
  0.9508 &
  0.0502 &
  0.000 &
  0.9874 &
  \multicolumn{1}{l}{0.0094} &
  0.1680 &
  - &
  - &
  - \\
 &
  Both &
  0.0866 &
  0.9151 &
  0.0866 &
  0.000 &
  0.9650 &
  \multicolumn{1}{l}{0.0339} &
  0.0900 &
  0.9851 &
  \multicolumn{1}{l}{0.0113} &
  0.1900 \\ \hline
\end{tabular}
}
\end{table}

\vspace{-1.2cm}
\begin{table}[htb]
\caption{Observed EER, accuracy, FAR, and FRR for the best blow-acoustic and face recognition techniques, and after combining them using score-level fusion.}
\label{tab:fusion_results}
\centering
\scalebox{0.85}{
\begin{tabular}{c|c|ccc|ccc|ccc}
\multirow{2}{*}{\begin{tabular}[c]{@{}c@{}}Biometrics\\ Features\end{tabular}} &
  \multirow{2}{*}{EER} &
  \multicolumn{3}{c|}{q = 10} &
  \multicolumn{3}{c|}{q = 9} &
  \multicolumn{3}{c}{q = 8} \\
                   &     & accuracy & FAR    & FRR   & accuracy & FAR    & FRR   & accuracy & FAR    & FRR   \\ \hline
blow-acoustic      & 0.0181 & 0.9822   & 0.0181 & 0.000 & 0.9924   & 0.0058 & 0.1000 & 0.9935   & 0.0027 & 0.192 \\ \hline
Facial Recognition & 0.0004 & 0.9996   & 0.0004 & 0.000 & 0.9980   & 0.0000 & 0.098 & 0.9960   & 0.0000 & 0.198 \\ \hline
Score-level fusion & 0.0018 & 0.9982   & 0.0018 & 0.000 & 0.9981   & 0.0000 & 0.094 & 0.9960   & 0.0000 & 0.198 \\ \hline
\end{tabular}
}
\end{table}

\vspace{-0.3cm}
\subsection{Usability Evaluation}

\subsubsection{User Convenience}
The proposed blow-acoustic technique is user-friendly and convenient for authentication. A simple blow on the phone enables the user to complete the authentication process. As such, it is an easy-to-use, intuitive, non-intrusive and highly practical behavioral biometric technique.
\subsubsection{Seamless Integration}

The proposed blow-acoustic behavioral biometric technique is 
seamlessly integrated with the facial recognition physiological method. Data from both modalities are captured simultaneously, and the authentication decision is made using a score-level fusion. Consequently, the blow-acoustic technique demonstrates strong compatibility for integration with other biometric modalities.

\subsection{Non-Invasiveness Evaluation}

The proposed blow-acoustic technique is entirely contactless and touch-free, requiring to perform only a simple blow directed at the phone screen. The procedure is also conducted within a matter of few seconds. Consequently, this technique is highly non-invasive and significantly reduces hygiene-related concerns and the risk of privacy violations associated with physical contact.

\subsection{Low Resource Requirements}

The proposed biometric technique does not require any additional or specialized hardware or software. It is compatible even with low-specification smartphones equipped solely with a camera and microphone. As such, it represents a highly cost-effective solution that is accessible to users with standard or budget-friendly mobile devices. 

\subsection{Comparison with Related Works}

In Table~\ref{table_comp_related_works}, we summarize the comparison between BlowPrint and the most closely related work discussed in Section~\ref{sec:related_work}. For each work, we present the best reported accuracy in terms of the metrics discussed in this section. Furthermore, we indicate whether they meet the other requirements discussed in Section~\ref{sec:requirements}: attack resistance, usability, non-invasiveness, MFB support, and low resource requirements. We observe that BlowPrint achieves the highest accuracy and is one of only three works to satisfy all requirements.

\begin{table}[htb]
\centering
\caption{Qualitative and quantitative comparison of related works}
\label{table_comp_related_works}
\scalebox{1}{%
\begin{tabular}{|l|ccccccccc|}
\hline
\multirow{3}{*}{Related work} &
  \multicolumn{9}{c|}{Evaluation Criteria} \\ \cline{2-10} 
 &
  \multicolumn{4}{c|}{EC1} &
  \multicolumn{1}{c|}{\multirow{2}{*}{EC2}} &
  \multicolumn{1}{c|}{\multirow{2}{*}{EC3}} &
  \multicolumn{1}{c|}{\multirow{2}{*}{EC4}} &
  \multicolumn{1}{c|}{\multirow{2}{*}{EC5}} &
  \multirow{2}{*}{EC6} \\ \cline{2-5}
 &
  \multicolumn{1}{c|}{Acc.} &
  \multicolumn{1}{c|}{FAR} &
  \multicolumn{1}{c|}{FRR} &
  \multicolumn{1}{c|}{EER} &
  \multicolumn{1}{c|}{} &
  \multicolumn{1}{c|}{} &
  \multicolumn{1}{c|}{} &
  \multicolumn{1}{c|}{} &
   \\ \hline
Chauhan et al. \cite{chauhan2017breathprint} &
  \multicolumn{1}{c|}{---} &
  \multicolumn{1}{c|}{$2\%$} &
  \multicolumn{1}{c|}{$6\%$} &
  \multicolumn{1}{c|}{---} &
  \multicolumn{1}{c|}{$\checkmark$} &
  \multicolumn{1}{c|}{$\checkmark$} &
  \multicolumn{1}{c|}{$\checkmark$} &
  \multicolumn{1}{c|}{$\times$} &
  $\checkmark$ \\ \hline
De Luca et al. \cite{de2012touch} &
  \multicolumn{1}{c|}{77\%} &
  \multicolumn{1}{c|}{21\%} &
  \multicolumn{1}{c|}{19\%} &
  \multicolumn{1}{c|}{---} &
  \multicolumn{1}{c|}{$\checkmark$} &
  \multicolumn{1}{c|}{$\checkmark$} &
  \multicolumn{1}{c|}{$\times$} &
  \multicolumn{1}{c|}{$\times$} &
  $\checkmark$ \\ \hline
Zheng et al. \cite{zheng2014you} &
  \multicolumn{1}{c|}{---} &
  \multicolumn{1}{c|}{---} &
  \multicolumn{1}{c|}{---} &
  \multicolumn{1}{c|}{3.65\%} &
  \multicolumn{1}{c|}{$\checkmark$} &
  \multicolumn{1}{c|}{$\checkmark$} &
  \multicolumn{1}{c|}{$\times$} &
  \multicolumn{1}{c|}{$\times$} &
  $\checkmark$ \\ \hline
Al-Waisy et al. \cite{al2017multimodal} &
  \multicolumn{1}{c|}{99\%} &
  \multicolumn{1}{c|}{---} &
  \multicolumn{1}{c|}{---} &
  \multicolumn{1}{c|}{---} &
  \multicolumn{1}{c|}{$\times$} &
  \multicolumn{1}{c|}{$\checkmark$} &
  \multicolumn{1}{c|}{$\checkmark$} &
  \multicolumn{1}{c|}{$\checkmark$} &
  $\checkmark$ \\ \hline
Aizi et al. \cite{aizi2022score} &
  \multicolumn{1}{c|}{95\%} &
  \multicolumn{1}{c|}{3.89\%} &
  \multicolumn{1}{c|}{1.5\%} &
  \multicolumn{1}{c|}{---} &
  \multicolumn{1}{c|}{$\times$} &
  \multicolumn{1}{c|}{$\checkmark$} &
  \multicolumn{1}{c|}{$\times$} &
  \multicolumn{1}{c|}{$\checkmark$} &
  $\checkmark$ \\ \hline
Srivastava et al. \cite{srivastava2022match} &
  \multicolumn{1}{c|}{99.7\%} &
  \multicolumn{1}{c|}{---} &
  \multicolumn{1}{c|}{---} &
  \multicolumn{1}{c|}{20\%} &
  \multicolumn{1}{c|}{$\times$} &
  \multicolumn{1}{c|}{$\times$} &
  \multicolumn{1}{c|}{$\times$} &
  \multicolumn{1}{c|}{$\checkmark$} &
  $\times$ \\ \hline
Mahfouz et al. \cite{mahfouz2024m2auth} &
  \multicolumn{1}{c|}{---} &
  \multicolumn{1}{c|}{---} &
  \multicolumn{1}{c|}{---} &
  \multicolumn{1}{c|}{0.84\%} &
  \multicolumn{1}{c|}{$\times$} &
  \multicolumn{1}{c|}{$\checkmark$} &
  \multicolumn{1}{c|}{$\times$} &
  \multicolumn{1}{c|}{$\checkmark$} &
  $\checkmark$ \\ \hline
El Rahman et al. \cite{el_rahman2020multimodal} &
  \multicolumn{1}{c|}{---} &
  \multicolumn{1}{c|}{---} &
  \multicolumn{1}{c|}{---} &
  \multicolumn{1}{c|}{---} &
  \multicolumn{1}{c|}{$\checkmark$} &
  \multicolumn{1}{c|}{$\checkmark$} &
  \multicolumn{1}{c|}{$\times$} &
  \multicolumn{1}{c|}{$\checkmark$} &
  $\times$ \\ \hline
Lee et al. \cite{lee2021advanced} &
  \multicolumn{1}{c|}{83\%} &
  \multicolumn{1}{c|}{1--7\%} &
  \multicolumn{1}{c|}{---} &
  \multicolumn{1}{c|}{---} &
  \multicolumn{1}{c|}{$\checkmark$} &
  \multicolumn{1}{c|}{$\checkmark$} &
  \multicolumn{1}{c|}{$\times$} &
  \multicolumn{1}{c|}{$\checkmark$} &
  $\times$ \\ \hline
Wu et al. \cite{wu2022echohand} &
  \multicolumn{1}{c|}{---} &
  \multicolumn{1}{c|}{---} &
  \multicolumn{1}{c|}{---} &
  \multicolumn{1}{c|}{2--3\%} &
  \multicolumn{1}{c|}{$\checkmark$} &
  \multicolumn{1}{c|}{$\checkmark$} &
  \multicolumn{1}{c|}{$\checkmark$} &
  \multicolumn{1}{c|}{$\checkmark$} &
  $\checkmark$ \\ \hline
Zhou et al. \cite{zhou2018echoprint} &
  \multicolumn{1}{c|}{93.75\%} &
  \multicolumn{1}{c|}{---} &
  \multicolumn{1}{c|}{10\%} &
  \multicolumn{1}{c|}{---} &
  \multicolumn{1}{c|}{$\checkmark$} &
  \multicolumn{1}{c|}{$\checkmark$} &
  \multicolumn{1}{c|}{$\checkmark$} &
  \multicolumn{1}{c|}{$\checkmark$} &
  $\checkmark$ \\ \hline
Ours &
  \multicolumn{1}{c|}{99.82\%} &
  \multicolumn{1}{c|}{0.18\%} &
  \multicolumn{1}{c|}{0\%} &
  \multicolumn{1}{c|}{0.18\%} &
  \multicolumn{1}{c|}{$\checkmark$} &
  \multicolumn{1}{c|}{$\checkmark$} &
  \multicolumn{1}{c|}{$\checkmark$} &
  \multicolumn{1}{c|}{$\checkmark$} &
  $\checkmark$ \\ \hline
\end{tabular}
}
\smallskip\\
\begin{tabularx}{\textwidth}{X}
\textbf{Description of notations:} 
EC1: Accuracy, EC2: Resilience against known attacks, EC3: Usability, EC4: Non-Invasiveness, EC5: MFB Support, EC6: Low Resource Requirements. 
$\checkmark$: Achieved, $\times$: Not Achieved, ---: Not Reported
\end{tabularx}
\end{table}

\section{Conclusion}
This paper presents a novel behavioral biometric authentication technique, called BlowPrint, that utilizes the phone blowing behavior of users to uniquely identify or authenticate them. To enhance its robustness and reliability, the technique was also seamlessly integrated with a facial recognition-based physiological biometric technique, forming a more effective multi-factor biometrics for a more secure and convenient user authentication.  

We employed various distance similarity algorithms alongside a machine learning-based classifier to compute 
the similarity scores of different blow-acoustic 
patterns, while we used a pretrained FaceNet-512 model for facial recognition. 
Subsequently, the similarity scores of the two modalities were combined using score-level fusion and the overall accuracy was computed using 
the $k$NN algorithm. The experimental results demonstrate that the proposed biometric technique offers high accuracy when compared to related works. Furthermore, it demonstrates several other advantages, such as high usability, non-invasiveness, and resilience against known attacks (e.g., spoofing attack) with minimal resource requirements.

The proposed protocol has significant applications in user authentication for online banking, smartphone unlocking, access control systems, and more. Future work will focus on improving the system’s robustness and scalability in real-world applications, particularly in dynamic environments. In addition, its robustness and resilience can be further improved by leveraging machine learning-based techniques.  
\section*{Acknowledgement}
This research is supported by the National Research Foundation, Singapore and Infocomm Media Development Authority under its Trust Tech Funding Initiative (DTC-T2FI-CFP-0002). Any opinions, findings and conclusions or recommendations expressed in this material are those of the author(s) and do not reflect the views of National Research Foundation, Singapore and Infocomm Media Development Authority.

\bibliographystyle{splncs04}
\bibliography{10_references}

\end{document}